\documentclass[english,12pt]{article}
\usepackage{array}
 \usepackage[normalem]{ulem}
\usepackage{graphicx}
\usepackage{amssymb}
\usepackage[english]{babel}
\usepackage{amsmath}
\usepackage{multirow}
\usepackage{prettyref}
\usepackage{babel}
\usepackage{units}
\usepackage[latin1]{inputenc}
\usepackage{amsfonts}
\usepackage{amssymb}
\usepackage{babel}
\usepackage{color}
\usepackage[dvipsnames]{xcolor} 
\usepackage{cite}
\def\@fmsl@sh#1#2#3{\m@th\ooalign{$\hfil#1\mkern#2/\hfil$\crcr$#1#3$}}
 \def\eq#1\en{\begin{equation}#1\end{equation}}
\def\s[#1,#2]{[#1\stackrel{\star}{,}#2]}
\def\sx[#1,#2]{[#1\stackrel{\star_{x}}{,}#2]}

\usepackage[english]{babel}

\usepackage{setspace}

\usepackage{graphicx} 
\usepackage{graphics} 
\usepackage{float} 
\usepackage{subcaption} 

\usepackage{amsmath}
\usepackage{amssymb}
\usepackage{mathtools}
\usepackage{enumerate}
\usepackage{enumitem}
\usepackage[colorlinks=true, allcolors=black]{hyperref}

\usepackage{tensor}

\newcommand\numberthis{\addtocounter{equation}{1}\tag{\theequation}}

\DeclareMathOperator\diag{diag}

\newcommand{\logbox}{\log\biggl(\frac{\Box}{\mu^2}\biggr)}
\newcommand{\logboxflat}{\log(\Box/\mu^2)}
\newcommand{\bM}{\bar{M}}
\newcommand{\bQ}{\bar{Q}}

\def\gsim{\mathrel{\rlap{\lower4pt\hbox{\hskip1pt$\sim$}}
		\raise1pt\hbox{$>$}}}       

\newcommand{\nc}{\newcommand}
\nc{\beq}{\begin{equation}}
\nc{\eeq}{\end{equation}}
\nc{\beqa}{\begin{eqnarray}}
\nc{\eeqa}{\end{eqnarray}}

\def\bc{\begin{center}}
\def\ec{\end{center}}

\def\to{\rightarrow}

\def\gsim{\mathrel{\mathpalette\atversim>}}

\def\bc{\begin{center}}
\def\ec{\end{center}}

\def\gsim{\mathrel{\rlap{\lower4pt\hbox{\hskip1pt$\sim$}}

    \raise1pt\hbox{$>$}}}       

\def\gsim{\mathrel{\rlap{\lower4pt\hbox{\hskip1pt$\sim$}}
    \raise1pt\hbox{$>$}}}       

\begin{document}
\makeatletter
\def\fmslash{\@ifnextchar[{\fmsl@sh}{\fmsl@sh[0mu]}}
\def\fmsl@sh[#1]#2{%
  \mathchoice
    {\@fmsl@sh\displaystyle{#1}{#2}}%
    {\@fmsl@sh\textstyle{#1}{#2}}%
    {\@fmsl@sh\scriptstyle{#1}{#2}}%
    {\@fmsl@sh\scriptscriptstyle{#1}{#2}}}
\def\@fmsl@sh#1#2#3{\m@th\ooalign{$\hfil#1\mkern#2/\hfil$\crcr$#1#3$}}

\makeatother

\thispagestyle{empty}
\begin{titlepage}
\boldmath
\begin{center}
  \Large {\bf {The Weak Gravity Conjecture in the Vilkovisky-DeWitt Effective Action of Quantum Gravity}}
    \end{center}
\unboldmath
\vspace{0.2cm}
\begin{center}
{\large Tommaso Antonelli}\footnote{t.antonelli@sussex.ac.uk}
 {\large and}  
{\large Xavier Calmet}\footnote{x.calmet@sussex.ac.uk}
\end{center}
\begin{center}
{\sl Department of Physics and Astronomy, 
University of Sussex, \\  Brighton, BN1 9QH, United Kingdom
}
\end{center}
\vspace{5cm}
\begin{abstract}
\noindent
The Weak Gravity Conjecture states that in any consistent theory of quantum gravity in the landscape of string theory, the repulsive force mediated by a U(1) gauge field must be stronger than the attractive force of gravity. In this work, we calculate quantum gravitational corrections to the charge-to-mass ratio of extremal Reissner-Nordstr\"om black holes, using the Vilkovisky-deWitt unique effective action approach, since this ratio provides a discriminant to select low-energy effective theories of quantum gravity that fulfill the conjecture. We find that depending on the values of the Wilson coefficients of the effective action, some UV complete theories will fulfill the conjecture, while others will not.
\end{abstract}  
\end{titlepage}



\newpage
\section{Introduction}
It is a well known fact that the electromagnetic repulsion between electrons is much stronger than their gravitational attraction by around 43 orders of magnitude. In other words, gravity is much weaker than electromagnetism. The same applies to the weak and strong interactions. Gravity is the weakest of the known forces.  Could gravity be weaker than any other conceivable force of nature including and in particular of any new force not yet discovered? 
The Weak Gravity Conjecture \cite{Arkani-Hamed:2006emk} (see \cite{Rudelius:2024mhq}, for a recent review) promotes this hypothesis  to a universal principle: any repulsive force, in any mathematically consistent universe, must be stronger than gravity, in the sense that the charge of a particle is larger than its mass, suitably normalized by the constants $k$ and $G$ where $G$ is Newton constant and $k$ some new constant of nature.  This question is particularly important in the context of the landscape of string theory as it could prove to be a useful criteria to discriminate between effective field theories that have a chance to be a consistent limit of string theory. 

A precise definition \cite{Rudelius:2024mhq} of the conjecture  is the following: given some effective field theory in the landscape with a U(1) gauge force, there exists a ``superextremal" particle, i.e., a particle that satisfies $q/m \geq Q/M|_{\mbox{ext}}$ where $Q/M|_{\mbox{ext}}$ is the charge-to-mass ratio of a large, extremal black hole.

This principle can be probed by studying the ratio charge over mass for an extremal Reissner-Nordstr\" om black hole. The charge-to-mass ratio of a finite-sized black hole can, in general, depend on the size of the black hole \cite{Kats:2006xp}:
\begin{eqnarray} \label{QoverM}
    \frac{Q}{M}\Bigg\rvert^{\mbox{finite M}}_{\mbox{ext}}=1-\epsilon(M)
\end{eqnarray}
where $|\epsilon(M)| \ll 1$ is a function of the mass $M$ of the extremal black hole satisfying
\begin{eqnarray}
\lim_{M\to \infty}\epsilon(M) = 0. 
\end{eqnarray}
An extremal black hole does not satisfy the weak gravity conjecture bound if $\epsilon(M)$ is negative. However, an ultraviolet complete theory could satisfy the Weak Gravity Conjecture  even if the quantity $\epsilon$ defined in Eq.(\ref{QoverM}) is negative. Indeed, it would not be satisfied by large, extremal black holes in this case, but it could still be satisfied by a light superextremal particle. However, if $\epsilon$ is negative, then the theory would violate the much stronger  Tower and the Sublattice Weak Gravity Conjectures as discussed in \cite{Harlow:2022ich}.

The aim of this Letter is to show that the ratio $Q/M|_{\mbox{ext}}$ for an extremal Reissner-Nordstr\" om black hole receives quantum gravitational corrections which can be reliably calculated using the Vilkovisky-deWitt unique effective action approach to quantum gravity \cite{Barvinsky:1983vpp,Barvinsky:1985an,Barvinsky:1987uw,Barvinsky:1990up,Buchbinder:1992rb,Calmet:2018elv}. We show that different theories of quantum gravity can have different values for this ratio. 

The unique effective action approach is obtained by integrating out the fluctuations of the graviton leading to higher-order curvature operators which have to be added to the Einstein-Hilbert action. The field equations derived from the unique effective action can be solved in perturbation theory around classical solutions of general relativity leading to quantum gravitational corrections to classical metric solutions. Some of the quantum gravitational corrections to the Reissner-Nordstr\"om metric at second order in curvature in the presence of these terms have already been partially calculated in the Refs.\cite{Delgado:2022pcc,amslaurea33432}, but as pointed out in \cite{Kats:2006xp}, when considering an extremal black hole more operators become relevant in a curvature expansion. We note that our work follows closely that of \cite{Kats:2006xp} which however only studied the local part of the effective action and it did not consider the non-local part of the effective action as it was not working in the same theoretical framework. The aim of the present work is to present a calculation of the quantum graviational corrections to the ratio $M/|Q|$ in the presence of both local and non-local terms in the unique quantum gravitational action.

\section{Corrections to the Reissner-Nordstr\"om geometry}

We start by establishing our notations. We consider the Reissner-Nordstr\"om metric arising from the equations of motion of General Relativity
\begin{equation}
  \label{eq:classical_metric}
  ds^2=f(r)\,dt^2-f(r)^{-1}\,dr^2-r^2d\Omega_{(2)}^2,\quad f(r)=1-\frac{2\kappa \bM}{r}+\frac{\kappa \bQ^2}{r^2}.
\end{equation}
Note that we are using the following conventions:
\begin{equation}
  c=\hbar=\varepsilon_0=\mu_0=1,\quad \kappa=4\pi G,\quad \bM=\frac{M}{4\pi},\quad \bQ=\frac{Q}{4\pi},
\end{equation}
$M$ and $Q$ being the black hole mass and  charge respectively. In this solution, the curvature tensors are given by
\begin{equation}
  \tensor{R}{^\mu_\nu}=\frac{\kappa \bQ^2}{r^4}\diag\{+1,+1,-1,-1\},\quad R=0
\end{equation}
and the non-zero components of the electromagnetic field strength tensor are
\begin{equation}
  \label{eq:classical_em_tensor}
  F_{01}=-F_{10}=\frac{\bQ}{r^2}.
\end{equation}
The location of the event horizon is given by
\begin{equation}
  f(r)=0\quad\implies\quad r_{\pm}=\kappa\bM\pm\left(\kappa^2\bM^2-\kappa\bQ^2\right)^{1/2}.
\end{equation}
As emphasized in the introduction, we consider extremal black holes, which are defined by the following relation
\begin{equation}
  \label{eq:classical_extremal_bh}
  r_{+}=r_{-} \quad\implies\quad r_{+}=\kappa\bM=\kappa^{1/2}\,|\bar Q|.
\end{equation}
The unique effective action formalism can only be applied for black holes with masses at least an order of magnitude the Planck mass \cite{Calmet:2021lny}. We thus work in the limit 
\begin{equation}
  \label{eq:mass_limit}
  \bM\gg M_P=G^{-1/2}.
\end{equation}

Now considering quantum corrections to the Einstein-Hilbert action, we are given a total effective action that can be written in the following form:
\begin{equation}
  \Gamma_{\text{tot}}=\Gamma_{\text{cl}}+\Gamma_{\text{L}}+\Gamma_{\text{NL}}.
\end{equation}
The local part is given by the classical Einstein-Hilbert
\begin{equation}
  \Gamma_{\text{cl}}=\int d^4x\,\sqrt{-g}\left[-\frac{1}{4\kappa}R-\frac{1}{4}F_{\mu\nu}F^{\mu\nu}\right]
\end{equation}
and by the higher-order operators
\begin{align*}
  \label{eq:final_operators}
  \Gamma_\text{L}=\int d^4x\,\sqrt{-g} &\Bigl[c_1 R_{\mu\nu}R^{\mu\nu}+c_2(F_{\mu\nu}F^{\mu\nu})^2 +c_3 R F_{\mu\nu}F^{\mu\nu}+\\
  &\ \ +c_4 R_{\mu\nu}F^{\mu\rho}\tensor{F}{^\nu_\rho}+c_5 R_{\mu\nu\rho\sigma}F^{\mu\nu}F^{\rho\sigma} +O\left(\bM^{-6}\right)\Bigr].  \numberthis
\end{align*}
The non-local part is given by
\begin{align*}
  \label{eq:final_operators_nl}
  \Gamma_{\text{NL}}=\int d^4x\,\sqrt{-g} \biggl[&b_1 R_{\mu\nu}\logbox R^{\mu\nu}+b_2F_{\mu\nu}F^{\mu\nu}\logbox F_{\rho\sigma}F^{\rho\sigma}+\\
  &+b_3 R\logbox F_{\mu\nu} F^{\mu\nu}+b_4 R_{\mu\nu}\logbox F^{\mu\rho}\tensor{F}{^\nu_\rho}+\\
  &+b_5 R_{\mu\nu\rho\sigma}\logbox F^{\mu\nu}F^{\rho\sigma}+O\left(\bM^{-6}\right)\biggr]. \numberthis
\end{align*}
The choice of the operators included in the effective action is explained in Appendix \ref{chap:app_A}. Note that the selection of the nonlocal operators follows from that of the local ones as they are related. In the Vilkovisky-DeWitt Effective Action framework, the non-local operators contain the finite parts of the loops diagrams under consideration while their local counterparts are used to absorb the divergent parts of the same diagrams. Local and nonlocal operators thus always come in pairs. Including both the local operators and the corresponding nonlocal one in any calculation is crucial to insure that the quantum corrections are renormalization group invariant.

Varying the action, we get that the equations of motion are
\begin{equation}
  \label{eq:tensor_EOMs}
  \begin{dcases}
    \tensor{G}{^\mu_\nu}=2\kappa \left(\tensor{T}{_{\text{cl}}^\mu_\nu}+\tensor{T}{_{\text{eff}}^\mu_\nu}\right)\\
    \nabla_{\nu}F^{\mu\nu}=\tensor{J}{_{\text{eff}}^\mu}
  \end{dcases}
\end{equation}
where $\tensor{T}{_{\text{cl}}^\mu_\nu}$ is the classical stress-energy tensor of $\Gamma_\text{cl}$, and where $\tensor{T}{_{\text{eff}}^\mu_\nu}$ and $\tensor{J}{_{\text{eff}}^\mu}$ are the effective stress-energy tensor and effective current respectively, derived from $\Gamma_{\text{L}}$ and $\Gamma_{\text{NL}}$. For the explicit form of these tensors consult Appendix \ref{chap:app_B}.

The solution to the equations (\ref{eq:tensor_EOMs}) can be given as a perturbation of the classical metric in (\ref{eq:classical_metric}):
\begin{equation}
  ds^2=e^{2\psi(r)}f(r)\,dt^2-f(r)^{-1}\,dr^2-r^2d\Omega_{(2)}^2, \quad f(r)=1-\frac{2\kappa\bM}{r}+\frac{\kappa\bQ^2}{r^2}-\frac{2\kappa \,m(r)}{r},
\end{equation}
and as a perturbation of the classical electromagnetic field strength tensor in (\ref{eq:classical_em_tensor}):
\begin{equation}
  F_{01}=-F_{10}=\frac{\bQ}{r^2}+F(r).
\end{equation}
These perturbations respect the spherical symmetry and staticity of the original solution. The unknown functions here are $m(r)$, $\psi(r)$ and $F(r)$, and we will solve the equations (\ref{eq:tensor_EOMs}) to leading order. Furthermore, we treat $\tensor{T}{_{\text{eff}}^\mu_\nu}$ and $\tensor{J}{_{\text{eff}}^\mu}$ as non-dynamical, so in (\ref{eq:tensor_EOMs}) they are simply evaluated at the background value of classical Reissner-Nordstr\"om. Note that the $\logboxflat$ distribution acts on both the curvature and the electromagnetic tensors in the way given in \cite{Calmet:Logbox}.

All these simplifications amount to some 
 simple coupled differential equations:
\begin{equation}
  \label{eq:differential_eq}
  \begin{dcases}
    m'(r)=\bQ F(r)+\frac{\bQ^2}{r^2}\psi(r)+A(r)\\
    F'(r)+\frac{\bQ}{r^2}\psi'(r)+\frac{2}{r}F(r)=B(r)
  \end{dcases}
\end{equation}
where $A(r)$ and $B(r)$ are known functions coming from $\tensor{T}{_{\text{eff}}^\mu_\nu}$ and $\tensor{J}{_{\text{eff}}^\mu}$. See again Appendix \ref{chap:app_B} for the explicit form of these functions.

The equations in (\ref{eq:differential_eq}) can be readily integrated for $m(r)$ as:
\begin{equation}
  m(r)=\int dr \,A(r)+\int dr\,\frac{1}{r^2}\int dr \,r^2\,\bQ B(r),
\end{equation}
where every constant of integration can be absorbed into $\bM$ and $\bQ$.

In the end, the form of the quantum-corrected metric factor $f(r)$ is given by
\begin{align*}
  f(r)=\ &1-\frac{2\kappa\bM}{r}+\frac{\kappa\bQ^2}{r^2}+\\
  &-\frac{2\kappa\bQ^2}{5r^6}\Bigg[4\,c_1\kappa\left(10r^2-15\kappa\bM r+6\kappa\bQ^2\right)+4\,c_2\bQ^2-20\,c_3\left(4r^2-7\kappa\bM r+3\kappa\bQ^2\right)-\\
  &\hspace{1.7cm}-2\,c_4\left(15r^2-25\kappa\bM r+11\kappa\bQ^2\right)-2\,c_5\left(20r^2-35\kappa\bM r+16\kappa\bQ^2\right)-\\
  &\hspace{1.7cm}-\frac{4}{5}\,b_1\kappa\left(50(L-4)r^2-25(3L-13)\kappa\bM r+6(5L-23)\kappa\bQ^2\right)-\\
  &\hspace{1.7cm}-\frac{4}{5}\,b_2(5L-13)\bQ^2+\\
  &\hspace{1.7cm}+20\,b_3\left(2(2L-7)r^2-(7L-25)\kappa\bM r+(3L-11)\kappa\bQ^2\right)+\\
  &\hspace{1.7cm}+\frac{2}{5}\,b_4\left(25(3L-11)r^2-25(5L-19)\kappa\bM r+(55L-213)\kappa\bQ^2\right)+\\
  &\hspace{1.7cm}+\frac{1}{5}\,b_5\left(200(L-4)r^2-25(14L-53)\kappa\bM r+16(10L-41)\kappa\bQ^2\right)\Bigg],\numberthis
\end{align*}
where $L=2\gamma+\log(\mu^2 r^2)$. These corrections, in the limit of a non-extremal black hole reduce to the ones calculated in \cite{Delgado:2022pcc,amslaurea33432}.       

Thus the leading correction to the weak gravity conjecture condition (\ref{eq:classical_extremal_bh}) is:
\begin{align*}
  \label{eq:final_ratio}
  \kappa^{1/2}\frac{\bM}{|\bQ|}\simeq1-\frac{2}{5\bQ^2}\bigg[&2\,c_1+2\,\frac{c_2}{\kappa^2}-\frac{c_4}{\kappa}-\frac{c_5}{\kappa}-\\
  &-\frac{2(5\bar{L}-13)}{5}\,b_1-\frac{2(5\bar{L}-13)}{5}\,\frac{b_2}{\kappa^2}+\\
  &+\frac{5\bar{L}-13}{5}\,\frac{b_4}{\kappa}+\frac{10\bar{L}-131}{10}\,\frac{b_5}{\kappa}\biggr],\numberthis
\end{align*}
where $\bar{L}=2\gamma+\log(\kappa\bQ^2\mu^2)$, which reproduces the results of \cite{Kats:2006xp} for the local part. Depending on the values of the Wilson coefficients, a UV complete theory of quantum gravity may or may not fulfill the weak gravity conjecture. This is thus a useful discriminant to disentangle between theories that are viable or not.  

\section{Conclusions}
Motivated by the weak gravity conjecture,  we have calculated quantum gravitational corrections to the ratio charge-over-mass of an extremal finite-sized Reissner-Nordstr\"om black hole using the Vilkovisky-deWitt effective action. Furthermore, we used an expansion in terms of the mass of the black hole, which is assumed to be some orders of magnitude larger than the Planck mass. This ensures the validity of the effective action. Note that our results can also be seen as an extension of the work of \cite{Kats:2006xp} who however, worked in a different framework for quantum gravity and thus did not use the same effective action.

We obtained the leading-order quantum gravitational corrections to ratio charge-over-mass and have verified that the result is renormalization group invariant.  The main result of our work is a quantitative calculation of the corrections to $\epsilon$ as a function of the Wilson coefficients of both local and nonlocal operators, as shown in Eq. (\ref{eq:final_ratio}). Depending on the values of the Wilson coefficients of the effective action, some UV complete theories will fulfill the conjecture, while others will not. We emphasize that while it is plausible that $\epsilon$ will always be non-negative for theories in the landscape, it has yet to be proven.

We have thus established a theoretical framework that enables one to discriminate between UV complete theories of quantum gravity using the weak gravity conjecture as a guiding principle.
Given a UV complete theory of quantum gravity, one can in principle integrate out the UV degrees of freedom and calculate the value of the Wilson coefficients $c_i$ in (\ref{eq:final_operators}) and $b_i$ in (\ref{eq:final_operators_nl}) (see e.g. \cite{Calmet:2024neu}), and then check whether the ratio in (\ref{eq:final_ratio}) is larger or smaller than one. 
\\
\\
{\it Acknowledgments:}
	The work of X.C. is supported in part  by the Science and Technology Facilities Council (grants numbers ST/T006048/1 and ST/Y004418/1.). The work of T.A. is supported by a doctoral studentship of the Science and Technology Facilities Council (training grant No. ST/Y509620/1, project ref. 2917813).
\\
\\
{\it Data Availability Statement:}
	This manuscript has no associated data. Data sharing not applicable to this article as no datasets were generated or analysed during the current study.
	
\newpage
\appendix
\section{Appendix - Selection of relevant operators for the effective action}
\label{chap:app_A}

We will follow the same reasoning as in \cite{Kats:2006xp}. As we are interested in the corrections to the ratio charge over mass of an extremal Reissner-Nordstr\"om black hole, we have to impose the condition (\ref{eq:classical_extremal_bh}). The radius $r$ is of the order of the event horizon $r_+$, and every partial derivative is of the order of $r_+^{-1}$. Then we list every possible term allowed by symmetries, expanding in powers of $\bM$, since we are assuming (\ref{eq:mass_limit}):
\begin{align*}
  \label{eq:initial_expansion_action}
  \Gamma=\Gamma_{\text{cl}}+\int d^4x\,\sqrt{-g} \Bigl[&\hat{c}_1 R^2+\hat{c}_2 R_{\mu\nu}R^{\mu\nu}+\hat{c}_3 R_{\mu\nu\rho\sigma}R^{\mu\nu\rho\sigma}+\hat{c}_4(F_{\mu\nu}F^{\mu\nu})^2+\\
  &+\hat{c}_5(F_{\mu\nu}\tilde{F}^{\mu\nu})^2+\hat{c}_6 \tensor{F}{^\mu_\nu}\tensor{F}{^\nu_\rho}\tensor{F}{^\rho_\sigma}\tensor{F}{^\sigma_\mu}+\hat{c}_7 \nabla_\mu F_{\nu\rho}\nabla^{\mu}F^{\nu\rho}+\\
  &+\hat{c}_8 \nabla_\mu F_{\nu\rho}\nabla^{\nu}F^{\mu\rho}+\hat{c}_9\nabla_\mu F^{\mu\rho}\nabla^{\nu}F_{\nu\rho}+\hat{c}_{10} R F_{\mu\nu}F^{\mu\nu}+\\
  &+\hat{c}_{11}R_{\mu\nu}F^{\mu\rho}\tensor{F}{^\nu_\rho}+\hat{c}_{12}R_{\mu\nu\rho\sigma}F^{\mu\nu}F^{\rho\sigma}+\hat{c}_{13}R_{\mu\nu\rho\sigma}F^{\mu\rho}F^{\nu\sigma}+
  \\&+O\left(\bM^{-6}\right)\Bigr]. \numberthis
\end{align*}
The quantum corrections to $\Gamma_{\text{cl}}$ will be treated in perturbation theory, and therefore only evaluated on the background values given by (\ref{eq:classical_metric}) and (\ref{eq:classical_em_tensor}).

This approximation allows us to drop some terms from the expansion in (\ref{eq:initial_expansion_action}). For example, the term proportional to $\hat{c}_1$ vanishes because, when calculating the equations of motion, one obtains a contribution proportional to the Ricci scalar $R$, which vanishes for the Reissner-Nordstr\"om metric. A similar fate happens to the terms proportional to $\hat{c}_5$ and to $\hat{c}_9$, since $F_{\mu\nu}\tilde{F}^{\mu\nu}$ and $\nabla_\mu F^{\mu\nu}$ both vanish when evaluated for the background. As pointed out in \cite{Kats:2006xp}, the term proportional to $\hat{c}_6$ can be regarded as half of the one proportional to $\hat{c}_4$ since the tensor $F_{\mu\nu}$ in the background has only $F_{01}=-F_{10}$ as non-zero components:
\begin{equation}
  \int d^4x\,\sqrt{-g}\,\tensor{F}{^\mu_\nu}\tensor{F}{^\nu_\rho}\tensor{F}{^\rho_\sigma}\tensor{F}{^\sigma_\mu}\quad = \quad \int d^4x\,\sqrt{-g}\,\frac{1}{2}(F_{\mu\nu}F^{\mu\nu})^2.
\end{equation}  
This and the following equalities are always meant to be evaluated on-shell using the Reissner-Nostr\"om metric as a background and up to boundary terms. 
The term proportional to $\hat{c}_3$ can be rewritten using the Gauss-Bonnet identity in 4 dimensions, which allows to relate it to the terms proportional to $\hat{c}_1$ and $\hat{c}_2$, plus a topological boundary term, which we disregard since we are only interested in the equations of motion of the theory:
\begin{equation}
  \int d^4x\,\sqrt{-g}\,R_{\mu\nu\rho\sigma}R^{\mu\nu\rho\sigma}\quad = \quad \int d^4x\,\sqrt{-g}\,\left[4R_{\mu\nu}R^{\mu\nu}-R^2\right].
\end{equation}
The term proportional to $\hat{c}_7$  can be  rewritten as a multiple of the term proportional to $\hat{c}_8$, by using the Bianchi identity $\nabla_{[\mu}F_{\nu\rho]}=0$, as:
\begin{align*}
  \int d^4x\,\sqrt{-g}\,\nabla_\mu F_{\nu\rho}\nabla^{\mu}F^{\nu\rho}\quad = \quad \int d^4x\,\sqrt{-g}\,2\nabla_\mu F_{\nu\rho}\nabla^{\nu}F^{\mu\rho}.\numberthis
\end{align*}
In turn, the term proportional to $\hat{c}_8$  can be  rewritten in terms of the contributions proportional to $\hat{c}_9$, $\hat{c}_{11}$ and $\hat{c}_{13}$, after integrating by parts one covariant derivative, swapping the order in which these derivatives act on the tensor, and re-integrating by parts, as:
\begin{align*}
  \int d^4x\,\sqrt{-g}\,\nabla_\mu F_{\nu\rho}\nabla^{\nu}F^{\mu\rho}\quad = \quad \int d^4x\,\sqrt{-g}\,\biggl[&\nabla_\mu F^{\mu\rho}\nabla^{\nu}F_{\nu\rho}-\\
  &-R_{\mu\nu}F^{\mu\rho}\tensor{F}{^\nu_\rho}+\frac{1}{2}R_{\mu\nu\rho\sigma}F^{\mu\rho}F^{\nu\sigma}\biggr].\numberthis
\end{align*}
The term proportional to $\hat{c}_{13}$ can be related to the one proportional to $\hat{c}_{12}$ using the first Bianchi identity of the Riemann tensor, combined with its usual symmetry properties:
\begin{equation}
  \int d^4x\,\sqrt{-g}\,R_{\mu\nu\rho\sigma}F^{\mu\rho}F^{\nu\sigma}\quad = \quad \int d^4x\,\sqrt{-g}\,\frac{1}{2}R_{\mu\nu\rho\sigma}F^{\mu\nu}F^{\rho\sigma}. 
\end{equation}

After dropping all these terms, the remaining ones are given in (\ref{eq:final_operators}), upon suitable redefinition of the constants $\hat{c}_i$ into $c_i$. The action now needs to be complemented with the appropriate non-local terms, which are the counterparts of the local ones and can be seen to remove the renormalization scale dependence of the field equations.

\section{Appendix - Effective tensors}
\label{chap:app_B}
The effective stress-energy tensor in equation (\ref{eq:tensor_EOMs}) is:
\begin{align*}
  \tensor{T}{_{\text{eff}}^\mu_\nu}=&\ c_1\left(4R^{\mu\rho}R_{\nu\rho}-4\nabla^\rho\nabla^\mu R_{\nu\rho}+2\Box \tensor{R}{^\mu_\nu}+2\delta^\mu_\nu\nabla_\rho\nabla_\sigma R^{\rho\sigma}-\delta^\mu_\nu R^{\rho\sigma}R_{\rho\sigma}\right)+\\
  &+c_2\left(8F^{\mu\lambda}F_{\nu\lambda}F^{\rho\sigma}F_{\rho\sigma}-\delta^\mu_\nu (F^{\rho\sigma}F_{\rho\sigma})^2\right)+\\
  &+c_3\bigl(2\tensor{R}{^\mu_\nu}F^{\rho\sigma}F_{\rho\sigma}+4RF^{\mu\rho}F_{\nu\rho}-2\nabla^\mu\nabla_\nu(F^{\rho\sigma} F_{\rho\sigma})+2\delta^\mu_\nu\Box (F^{\rho\sigma} F_{\rho\sigma})-\\
  &\hspace{1.1cm}-\delta^\mu_\nu R F^{\rho\sigma}F_{\rho\sigma}\bigr)+\\
  &+c_4\bigl(4\tensor{R}{_\nu^\rho}F^{\mu\sigma}F_{\rho\sigma}+2\tensor{R}{_\rho^\sigma}F^{\mu\rho}F_{\nu\sigma}-2\nabla_\rho\nabla^\mu ( F^{\rho\sigma}F_{\nu\sigma})+\Box (F^{\mu\rho}F_{\nu\rho})+\\
  &\hspace{1.1cm}+\delta^\mu_\nu \nabla_\rho\nabla^\sigma(F^{\rho\lambda} F_{\sigma\lambda})-\delta^\mu_\nu \tensor{R}{_\rho^\sigma} F^{\rho\lambda}F_{\sigma\lambda}\bigr)+\\
  &+c_5\bigl(6\tensor{R}{_{\nu\rho}^{\sigma\lambda}}F^{\mu\rho}F_{\sigma\lambda}+4 \nabla_\rho\nabla^\sigma(F^{\mu\rho}  F_{\nu\sigma})-\delta^\mu_\nu \tensor{R}{_{\rho\sigma}^{\lambda\tau}}F^{\rho\sigma}F_{\lambda\tau}\bigr)+\\
  &+b_1\bigl(4\tensor{R}{_{\rho}^{\mu}}\logboxflat \tensor{R}{_{\nu}^{\rho}}-4\nabla_\rho\nabla^\mu \logboxflat \tensor{R}{_{\nu}^{\rho}}+2\Box \logboxflat\tensor{R}{_\nu^\mu}+\\
  &\hspace{1.1cm}+2\delta^\mu_\nu\nabla^\rho\nabla_\sigma \logboxflat \tensor{R}{_{\rho}^{\sigma}}-\delta^\mu_\nu \tensor{R}{_{\sigma}^{\rho}}\logboxflat \tensor{R}{_{\rho}^{\sigma}}\bigr)+\\
  &+b_2\bigl(8F^{\mu\lambda}F_{\nu\lambda}\logboxflat (F^{\rho\sigma}F_{\rho\sigma})-\delta^\mu_\nu F^{\rho\sigma}F_{\rho\sigma}\logboxflat (F^{\lambda\tau}F_{\lambda\tau})\bigr)+\\
  &+b_3\bigl(2\tensor{R}{^\mu_\nu}\logboxflat (F^{\rho\sigma}F_{\rho\sigma})+4R\logboxflat(F^{\mu\rho}F_{\nu\rho})-\\
  &\hspace{1.1cm}-2\nabla^\mu\nabla_\nu\logboxflat(F^{\rho\sigma} F_{\rho\sigma})+2\delta^\mu_\nu\Box \logboxflat(F^{\rho\sigma} F_{\rho\sigma})-\\
  &\hspace{1.1cm}-\delta^\mu_\nu R \logboxflat(F^{\rho\sigma}F_{\rho\sigma})\bigr)+\\
  &+b_4\bigl(4\tensor{R}{_\nu^\rho}\logboxflat(F^{\mu\sigma}F_{\rho\sigma})+2\tensor{R}{_\rho^\sigma}\logboxflat(F^{\mu\rho}F_{\nu\sigma})-\\
  &\hspace{1.1cm}-2\nabla_\rho\nabla^\mu\logboxflat (F^{\rho\sigma}F_{\nu\sigma})+\Box \logboxflat(F^{\mu\rho}F_{\nu\rho})+\\
  &\hspace{1.1cm}+\delta^\mu_\nu \nabla_\rho\nabla^\sigma\logboxflat(F^{\rho\lambda} F_{\sigma\lambda})-\delta^\mu_\nu \tensor{R}{_\rho^\sigma} \logboxflat(F^{\rho\lambda}F_{\sigma\lambda})\bigr)+\\
  &+b_5\bigl(6\tensor{R}{_{\nu\rho}^{\sigma\lambda}}\logboxflat(F^{\mu\rho}F_{\sigma\lambda})+4 \nabla_\rho\nabla^\sigma\logboxflat(F^{\mu\rho}  F_{\nu\sigma})-\\
  &\hspace{1.1cm}-\delta^\mu_\nu \tensor{R}{_{\rho\sigma}^{\lambda\tau}}\logboxflat(F^{\rho\sigma}F_{\lambda\tau})\bigr). \numberthis
\end{align*}
Meanwhile, the effective current in (\ref{eq:tensor_EOMs}) is given by:
\begin{align*}
  \tensor{J}{_{\text{eff}}^\mu}=&\ 8c_2\nabla_\nu\left(F^{\mu\nu}F_{\rho\sigma} F^{\rho\sigma}\right)+4c_3\nabla_\nu\left( F^{\mu\nu}R\right)+\\
  &+4c_4\nabla_\nu\left(F^{\rho[\nu}\tensor{R}{^{\mu]}_\rho}\right)+4c_5\nabla_\nu\left(F^{\rho\sigma}\tensor{R}{^{\mu\nu}_{\rho\sigma}}\right)+\\
  &+8b_2\nabla_\nu\left(F^{\mu\nu}\logboxflat (F_{\rho\sigma} F^{\rho\sigma})\right)+\\
  &+4b_3\nabla_\nu\left( F^{\mu\nu}\logboxflat R\right)+4b_4\nabla_\nu\left(F^{\rho[\nu}\logboxflat\tensor{R}{^{\mu]}_\rho}\right)+\\
  &+4b_5\nabla_\nu\left(F^{\rho\sigma}\logboxflat\tensor{R}{^{\mu\nu}_{\rho\sigma}}\right).\numberthis
\end{align*}
The functions $A(r)$ and $B(r)$ are given by:
\begin{align*}
  A(r)=&\,\frac{2\bQ^2}{r^4}(-12\kappa c_1+24c_3+9c_4+12c_5-\\
  &\hspace{1cm}-56\kappa b_1+12L\kappa b_1+100b_3-24Lb_3+39b_4-9Lb_4+56b_5-12Lb_5)+\\
  &+\frac{4\kappa\bM\bQ^2}{r^5}(12\kappa c_1-28c_3-10c_4-10c_5+\\
  &\hspace{2.1cm}+58\kappa b_1-12L\kappa b_1-114b_3+28Lb_3-43b_4+10Lb_4-52b_5+10Lb_5)+\\
  &+\frac{2\bQ^4}{r^6}(-12\kappa^2c_1+6c_2+30\kappa c_3+9\kappa c_4+4\kappa c_5-\\
  &\hspace{1.5cm}-60b_1\kappa^2+12L\kappa^2b_1+18b_2-6Lb_2+122\kappa b_3-30L\kappa b_3+41\kappa b_4-9L\kappa b_4+\\
  &\hspace{1.5cm}+36 \kappa b_5-4L\kappa b_5)\numberthis
\end{align*}
\begin{align*}
  B(r)=&-\frac{16\kappa\bM\bQ}{r^6}(-3c_5-8b_5+3Lb_5)+\\
  &+\frac{8\bQ^3}{r^7}(8c_2-2\kappa c_4-12\kappa c_5+28b_2-8Lb_2-7\kappa b_4+2L\kappa b_4-42\kappa b_5+12L\kappa b_5)\numberthis
\end{align*}
where again $L=2\gamma+\log(\mu^2r^2)$.


\bigskip{}


\baselineskip=1.6pt 

\begin{thebibliography}{10}

\bibitem{Arkani-Hamed:2006emk}
N.~Arkani-Hamed, L.~Motl, A.~Nicolis and C.~Vafa,
JHEP \textbf{06} (2007), 060
doi:10.1088/1126-6708/2007/06/060
[arXiv:hep-th/0601001 [hep-th]].
  
\bibitem{Rudelius:2024mhq}
T.~Rudelius,
Contemp. Phys. \textbf{1} (2024), 14
doi:10.1080/00107514.2024.2391206
[arXiv:2409.02161 [hep-th]].

\bibitem{Harlow:2022ich}
D.~Harlow, B.~Heidenreich, M.~Reece and T.~Rudelius,
Rev. Mod. Phys. \textbf{95} (2023) no.3, 3
doi:10.1103/RevModPhys.95.035003
[arXiv:2201.08380 [hep-th]].

\bibitem{Kats:2006xp}
Y.~Kats, L.~Motl and M.~Padi,
JHEP \textbf{12}, 068 (2007)
doi:10.1088/1126-6708/2007/12/068
[arXiv:hep-th/0606100 [hep-th]].


    \bibitem{Barvinsky:1983vpp}
    A.~O.~Barvinsky and G.~A.~Vilkovisky,
    Phys. Lett. B \textbf{131}, 313-318 (1983).
    
    \bibitem{Barvinsky:1985an} 
    A.~O.~Barvinsky and G.~A.~Vilkovisky,
    Phys.\ Rept.\  {\bf 119}, 1 (1985).
    
    \bibitem{Barvinsky:1987uw} 
    A.~O.~Barvinsky and G.~A.~Vilkovisky,
    Nucl.\ Phys.\ B {\bf 282}, 163 (1987).
    
    \bibitem{Barvinsky:1990up} 
    A.~O.~Barvinsky and G.~A.~Vilkovisky,
    Nucl.\ Phys.\ B {\bf 333}, 471 (1990).
    
    \bibitem{Buchbinder:1992rb} 
    I.~L.~Buchbinder, S.~D.~Odintsov and I.~L.~Shapiro,
    ``Effective action in quantum gravity,''
    (CRC Press, Bristol, 1992).
    
\bibitem{Calmet:2018elv}
X.~Calmet,
Phys. Lett. B \textbf{787} (2018), 36-38
doi:10.1016/j.physletb.2018.10.040
[arXiv:1810.09719 [hep-th]].


\bibitem{Delgado:2022pcc}
R.~C.~Delgado,
Eur. Phys. J. C \textbf{82}, no.3, 272 (2022)
[erratum: Eur. Phys. J. C \textbf{83}, no.6, 468 (2023)]
doi:10.1140/epjc/s10052-022-10232-0
[arXiv:2201.08293 [hep-th]].

\bibitem{amslaurea33432}
I.~Perrucci,
http://amslaurea.unibo.it/33432/


\bibitem{Calmet:2021lny}
X.~Calmet and F.~Kuipers,
Phys. Rev. D \textbf{104} (2021) no.6, 066012
doi:10.1103/PhysRevD.104.066012
[arXiv:2108.06824 [hep-th]].



\bibitem{Calmet:Logbox}
X.~Calmet, R.~Casadio and F.~Kuipers
Phys. Rev. D, \textbf{100} (2019) no.8, 086010
doi:10.1103/PhysRevD.100.086010
[arXiv:1909.13277 [hep-th]]



\bibitem{Calmet:2024neu}
X.~Calmet, E.~Kiritsis, F.~Kuipers and D.~Lust,
Fortsch. Phys. \textbf{2024}, 2400176
doi:10.1002/prop.202400176
[arXiv:2408.15146 [hep-th]].



\end{thebibliography}
\end{document}